**Title of the article:**

ECCOLA - a Method for Implementing Ethically Aligned AI Systems

**Authors:**

Ville Vakkuri, Kai-Kristian Kemell, Pekka Abrahamsson

**Notes:**

- This is the author's version of the work.
- ECCOLA method's Internet resources are accessible at: https://doi.org/10.6084/m9.figshare.12136308

- The definite version was published in: V. Vakkuri, K. -K. Kemell and P. Abrahamsson, "ECCOLA - a Method for Implementing Ethically Aligned AI Systems," 2020 46th Euromicro Conference on Software Engineering and Advanced Applications (SEAA), Portoroz, Slovenia, 2020, pp. 195-204, doi: 10.1109/SEAA51224.2020.00043.

- Copyright owner's version can be accessed at https://doi.org/10.1109/SEAA51224.2020.00043

# ECCOLA - a Method for Implementing Ethically Aligned AI Systems

Ville Vakkuri [0000-0002-1550-1110]
Faculty of Information Technology
University of Jyväskylä
Jyväskylä, Finland
ville.vakkuri@jyu.fi

Kai-Kristian Kemell [0000-0002-0225-4560]
Faculty of Information Technology
University of Jyväskylä
Jyväskylä, Finland
kai-kristian.o.kemell@jyu.fi

Pekka Abrahamsson [0000-0002-4360-2226]
Faculty of Information Technology
University of Jyväskylä
Jyväskylä, Finland
pekka.abrahamsson@jyu.fi

*Abstract*— Various recent Artificial Intelligence (AI) system failures, some of which have made the global headlines, have highlighted issues in these systems. These failures have resulted in calls for more ethical AI systems that better take into account their effects on various stakeholders. However, implementing AI ethics into practice is still an on-going challenge. High-level guidelines for doing so exist, devised by governments and private organizations alike, but lack practicality for developers. To address this issue, in this paper, we present a method for implementing AI ethics. The method, ECCOLA, has been iteratively developed using a cyclical action design research approach. The method aims at making the high-level AI ethics principles more practical, making it possible for developers to more easily implement them in practice.

1. INTRODUCTION

As we make increasing progress on Artificial Intelligence (AI), the systems become increasingly widespread and exert a growing impact on society. This has also resulted in us witnessing various AI system failures, which have served to highlight various ethical issues associated with these systems. Many of these failures have made the global headlines and resulted in public backlash. Especially privacy issues related to facial recognition technology have become a prominent topic among the general public, as well as for policymakers.

The systems we develop, despite us having had some collective learning experiences from past system failures, are still far from being problem-free. Ethical issues persist, and more arise as the technologies become more sophisticated. Aside from the obvious physical damage potential of systems such as autonomous vehicles, data handling alone is ripe with ethical issues without universal answers.

The discussion on the field of AI ethics has soared in activity in the past decade following this technological progress, resulting in the birth of some key principles that are now widely acknowledged as central issues in AI ethics. One such issue is the demand for AI systems that are explainable [1]. The problem thus far has been transferring this discussion into practice. I.e., how to actually influence the development of these systems?

For the time being, this has mostly been carried out either via guidelines or laws and regulations. Guidelines have been devised by companies [2], governments [3] and standardization organizations [4]. Yet, these guidelines have been lacking in actionability. Developers struggle to implement abstract ethical guidelines into the development process [5,6].

Methods and practices in the area remain highly technical, focusing on specific issues in e.g. machine learning [7]. While certainly useful in their specific contexts, these types of tools do not help companies in the design and development process as a whole. Thus, development methods are still required to bridge this gap between research and practice in the area.

In this paper, we present our work on an AI ethics method: ECCOLA. ECCOLA has been developed iteratively over the past two years through empirical use and data resulting from it, with each iteration improving the method. ECCOLA is intended to help organizations implement AI ethics in practice, in an actionable manner.

The rest of this paper is structured as follows. The second section discusses the theoretical background of the paper: AI ethics, methods in AI ethics, as well as the Essence Theory of Software Engineering used in devising the method in question. The third section presents the method, ECCOLA. In the fourth section we discuss how ECCOLA was iteratively developed and what kind of data were used in doing so. In the fifth and final section we discuss the method in relation to extant literature and conclude the paper.

2. THEORETICAL BACKGROUND

This section is split into three subchapters. In the first one, we provide an overview of the current state of AI ethics in research. In the second one, we focus on the state of the practical implementation of AI ethics, discussing the methods and other tools that currently exist to help practitioners implement it. In the third and final one, we discuss the Essence Theory of Software Engineering, and specifically the idea of essentializing software engineering practices, as this an approach we have utilized in devising ECCOLA.

A. AI Ethics

AI ethics is a long-standing area of research. In the past, much of the debate focused on hypothetical future scenarios that would result from technological progress.

However, as these hypothetical future scenarios start to become reality following said progress, which to many has been faster than anticipated, the field has become increasingly active.

Much of the research in the area has focused on theory, and specifically to define AI ethics by highlighting key ethical issues in AI systems. This discussion has focused on principles. Many have been proposed and discussed, and, by now, some have become largely agreed-upon [8]. Based on an analysis of the numerous AI ethics guidelines that now exist, Jobin et al. [8] listed the key principles that could be considered central based on how often they appear in these guidelines: "transparency, justice and fairness, non-maleficence, responsibility, privacy, beneficence, freedom and autonomy, trust, dignity, sustainability, and solidarity."

To provide an example of the type of research that has been conducted on these principles, we can look at transparency. Transparency [9] is widely considered one of the central AI ethical principles. Transparency is about understanding AI systems, how they work, and how they were developed [9,10]. It has been argued to be the very foundation of AI ethics: if we cannot understand how the systems work, we cannot make them ethical either [11]. The discussion on transparency has, aside from defining what it is, focused on how to achieve it. For example, Ananny & Crawford [10] discussed the limitations of the idea of transparency in relation to the complexity brought on by machine learning. Is being able to see inside the system really enough or even helpful? Transparency is featured as a key principle in the high-profile guidelines of EU [3] and IEEE [4], for example.

Though principles are one way of categorizing the discussion in the area, it is ultimately about bringing attention to potential ethical issues in AI, with or without pinning them under a specific principle. Privacy issues, for example, have been one prominent topic of discussion both in academia and the media following various practical examples of (ethical) AI system failures. Privacy issues have been discussed in relation to data handling, technology such as facial recognition, as well as racial bias, which falls under the principle of fairness.

Indeed, guidelines have, thus far, been the main way of bridging the gap between research and practice in the area. The purpose of these guidelines has been to distill the discussion in the area into a tool. However, past research has shown that guidelines are rarely effective in software engineering. McNamara et al. [6] studied the impact the ACM Code of Ethics[1] had had on practice in the area, finding little to none. This seems to also be the case in AI ethics: in a recent paper [5], we studied the current state of practice in AI ethics and found that the principles present in literature are not actively tackled out on the field.

This state of affairs underlines a need for more actionable tools for implementing AI ethics in practice. In the context of software engineering, we thus turn to methods, i.e. ways of working that direct how work is carried out [12]. As software engineering in any organization is carried out using typically some form of an agile method [13], hybrid or in-house ones, incorporating AI ethics into these methods would be a goal to strive for.

### B. Methods in AI Ethics

There are already various methods and tools for implementing AI ethics, as highlighted by Morley et al. [7] in their systematic review. These are largely tools for the technical side of AI system development, such as tools for machine learning. On the other hand, we are not currently aware of any method focusing on the higher-level design and development decisions surrounding AI systems. Guidelines have been devised for this but seem to remain impractical given their seeming lack of adoption out on the field [5].

Aside from AI ethics methods and tools, some ethical tools from other fields do exist that could potentially be used to design ethical AI systems. One example of such a tool is the RESOLVEDD method from the field of business ethics [14]. We have, in a past study [15], studied the suitability of this particular method for the AI ethics context, with our results suggesting that dedicated methods would be more beneficial. Such methods, however, are currently lacking.

Aside from ECCOLA, there is currently some other activity in method development for the area as well, though to the best of our knowledge most of these are still work-in-progress. E.g., while not a software engineering method as such, Leikas et al. [16] recently presented an "ethical framework for designing autonomous intelligent systems". This framework, however, is more focused on higher level design than development and not specifically aimed at developers or product managers.

In devising ECCOLA, our method, we have turned to the Essence Theory of Software Engineering for method engineering. Specifically, we have utilized the theory's philosophy of essentializing software engineering practices in devising a method, as we discuss next.

### C. Essentializing to Create Methods from Practices

The Essence Theory of Software Engineering (Jacobson et al. [12]) is a method engineering tool. It comprises a method core, which the authors refer to as a kernel, as well as a language. The kernel, they argue [12], contains all the core elements present in any software engineering project.

To this end, the kernel contains three types of items: alphas (ie. things to work with), activities (things to do), and competencies (skills required to carry out the tasks). There are seven alphas, which form the core of the kernel[2]: opportunity, stakeholders, requirements, software system, work, team and way-of-working. The kernel provides a basis for constructing methods using the Essence language to describe them. I.e., the theory consists of basic building blocks which can be utilized by using the language to extend the base to build a method. On its own, the kernel could be used as a generic software engineering method, but the point of Essence is to construct new methods using the language, while utilizing the kernel as an extensible starting point for doing so.

---

[1] https://www.acm.org/code-of-ethics

Software engineering methods consist of practices. A practice is a more atomic unit of work, such as pair programming. In creating ECCOLA, we have utilized the idea of essentializing [17] software engineering practices. In short, this refers to describing them using the Essence language. This offers one way of breaking down practices into different elements in order to describe them, making them easier to understand. This also serves to make practices more modular, as describing them in the same notational language makes it easier to combine them into methods.

Essentializing practices is described as a process by Jacobson [17] as follows:

> *"- Identifying the elements – this is primarily identifying a list of elements that make up a practice. The output is essentially a diagram [...]*
> *- Drafting the relationships between the elements and the outline of each element – At this point, the cards are created.*
> *- Providing further details – Usually, the cards will be supplemented with additional guidelines, hints and tips, examples, and references to other resources, such as articles and books"*

As can be observed in the above quote, Essence utilizes cards to describe methods. This is also an approach we have utilized in ECCOLA: ECCOLA is a card deck.

Essence was also chosen due to its method-agnostic approach and modular philosophy on methods. From the get-go, ECCOLA was never intended to be a stand-alone method, but rather, a modular extension to existing software development methods that would bring in AI ethics.

Originally, we planned on using the Essence language to describe ECCOLA. For example, principles such as transparency could have been alphas (i.e. things to work with) in the method. However, as the development of the method progressed and we began to test its early versions in practice, Essence turned out to make the method confusing to its users. As a result, the role of Essence in ECCOLA grew smaller, as we discuss in the fourth section.

3. ECCOLA - A METHOD FOR DESIGNING ETHICALLY ALIGNED AI SYSTEMS

As we have discussed in section II, AI ethics is currently an area with a prominent gap between research and practice. Much of the research has been theoretical and conceptual, focusing on defining key principles for AI ethics and how to tackle them. The numerous guidelines for AI ethics that currently exist [8] have tried to bridge this gap to bring these principles to the developers, but seem to not have had much success. Indeed, ethical guidelines tend to not have much impact in the context of SE [6]. To bridge this gap we propose a method for implementing AI ethics: ECCOLA.

ECCOLA[3] (figure 1) is intended to provide developers an actionable tool for implementing AI ethics. To utilize the various AI ethics guidelines in practice, the organization seeking to do so has to somehow make them practical first. ECCOLA, on the other hand, is intended to be practical as is, and ready to be incorporated into any existing method. ECCOLA does not provide any direct answers to ethical problems, as arguably correct answers are a rare breed in ethics in general, but rather asks questions in order to make the organization consider the various ethical issues present in AI systems. Though ultimately how these questions are then tackled is up to the organization in question, ECCOLA does encourage taking into account any ethical issues it highlights.

ECCOLA is built on AI ethics research. It utilizes both existing theoretical and conceptual research, as well as AI ethics guidelines that have been devised based on existing research as well. In terms of guidelines, the cards are based primarily on the IEEE Ethically Aligned Design guidelines [4] and the EU Trustworthy AI guidelines [4]. As these guidelines have already distilled much of the existing research on the topic under various principles, these principles have been utilized in ECCOLA as well. AI ethics research has been used to further expand on these principles in ECCOLA.

In practice, ECCOLA takes on a form of a deck of cards. This approach was based on the Essence Theory of Software Engineering [12], which was used to describe the first versions of the method. Methods described using the Essence language are utilized through cards. However, using cards in the context of software engineering methods is not a novel idea, nor one proposed by Essence. E.g., Planning Poker in Agile uses cards and the idea of Kanban is founded around using cards in the form of sticky notes.

There are 21 cards in total In ECCOLA. These cards are split into 8 themes, with each theme consisting of 1 to 6 cards. These themes are AI ethics themes found in various ethical guidelines [8], such as transparency or data. Each individual card, then, deals with a more atomic aspect of that theme, such as, in the case of data, data privacy and data quality. Aside from the main set of cards, ECCOLA also features an A5-sized game sheet that describes how the method is used.

---

[2] http://semat.org/alpha-definitions-overview/competency-cards

[3] https://figshare.com/articles/Internet_resource_for_ECCOLA_-_a_Method_for_Implementing_Ethically_Aligned_AI_Systems/12136308

# ECCOLA

## Game Sheet – How to Play the Cards

**Info:** ECCOLA is easy to apply in practice. It is a sprint-by-sprint evolving process that empowers ethical thinking in the product development process. As a result, ethical development is enhanced and Work Product Sheets (WPS) are created. The WPSs help you measure the Trustworthiness of the product. ECCOLA is an evolving set of cards and you choose the parts that are relevant to your work.

**How to:** ECCOLA is intended to be used during the entire design and development process in three steps:
1. Prepare – Choose the relevant cards for the current sprint. Document selected cards and justification on WPS.
2. Review – Keep the selected cards on hand during single tasks. Write down if any actions are taken based on the cards.
3. Evaluate – Review to ensure that all planned actions are taken. Revise the card deck, and if necessary, review tasks again.

**Practical Tip:** Repeat the process in every iteration. Remember to do a retrospective afterwards. Think about what worked & what did not. Choose the parts that are the most relevant for your work in the next round.

1. Prepare → 2. Review → 3. Evaluate

---

### #0 Stakeholder Analysis (Analyze)

**Motivation:** In order to understand the big picture, it is important to first understand who the system can affect and how. Try to also think past the obvious, direct stakeholders such as your end-users.

**What to Do:** Identify stakeholders.
- Who does the system affect, and how? Stakeholders are not simply users, developers and customers.
- How are the various stakeholders linked together?
- Can these different stakeholders influence the development of the system? How?
- Remember that a user is often an organization and the end-user is an individual. Similarly, AI systems can treat people as objects for data collection.

**Practical Example:** Autonomous cars don't just affect their passengers. Anyone nearby is affected; some even change the way they drive. If at one point half of the traffic consists of self-driving cars, what are the societal impacts of such systems? E.g., how are the people who can't afford one affected? Regulations arising from such systems also affect everyone.

---

### #1 Types of Transparency (Transparency)

**Motivation:** When considering transparency, it is important to understand who you are being transparent towards, and what you are being transparent about.

**What to Do:** Consider the following…
- Are you trying to understand something? (Internal transparency)
- Are you trying to explain something? (External transparency)
- Are you trying to understand or explain how the system works? (Transparency of algorithms and data)
- Are you trying to understand or explain why the system was made to be the way it now is? (Transparency of system development)
- External stakeholders to consider, among others: (end-)users, safety certification agencies, accident investigators, lawyers or expert witnesses, and society at large for disruptive technologies

---

### #2 Explainability (Transparency)

**Motivation:** If we cannot understand the reasons behind the actions of the AI, it is difficult to trust it.

**What to Do:** Ask yourself:
- Is explainability a goal for your system? How do you plan to ensure it?
- How well can each decision of the system be understood? By both developers and (end-)users.
- Did you try to use the simplest and most interpretable model possible for the context?
- Did you make trade-offs between explainability and accuracy? What kind of? Why?
- How familiar are you with your training or testing data? Can you change it when needed?
- If you utilize third party components in the system, how well do you understand them?

**Practical Example:** When interacting with a robot, users could ideally ask the robot "why did you do that?" and receive an understandable response. This would make it much easier for them to trust a system.

---

### #6 System Reliability (Transparency)

**Motivation:** Transparency makes ethical development possible in the first place. To make it ethical, we must understand how the system works and why it makes certain decisions.

**What to Do:** Ask yourself:
- How do you test if the system fulfills its goals?
- Have you tested the system comprehensively, including unlikely scenarios? Have the tests been documented?
- When the system fails in a certain scenario, will you be able to tell why? Can you replicate the failure?
- How do you assure the (end-)user of the system's reliability?

**Practical Example:** An autonomous coffee machine successfully brews coffee 8 times out 10. While this is a decent success rate, we are left wondering what happened the 2 times it failed to do so, and why. Errors are inevitable, but we must understand the causes behind them and be able to replicate them to fix them.

---

### #5 Traceability (Transparency)

**Motivation:** Traceability supports explainability. It helps us understand why the AI acts the way it does.

**What to Do:** Document. Different types of documentation (code, project etc.) are typically key in producing transparency.
- How have you documented the development of the system, both in terms of code and decision-making? How was the model built or the AI trained?
- How have you documented the testing and validation process? In terms of data and scenarios used etc.
- How do you document the actions of the system? What about alternate actions (e.g. if the user was different but the situation otherwise the same)?

**Practical Example:** When the system starts making mistakes, by aiming for traceability, it will be easier to find out the cause. Consequently, it will also be faster and possibly easier to start fixing the underlying issue.

---

### #4 Documenting Trade-offs (Transparency)

**Motivation:** One important part of transparent system development is the documentation of trade-offs. Whenever you make a decision, you choose one option over other alternatives. However, documenting why and what the alternatives were is important.

**What to Do:** Ask yourself:
- Are relevant interests and values implicated by the system and potential trade-offs between them identified and documented?
- Who decides on such trade-offs (e.g. between two competing solutions) and how? Did you ensure that the trade-off decision and the reasons behind it were documented?

**Practical Example:** Documenting trade-offs can improve your customer relationship, allowing you to better explain why certain decisions were made over others. Moreover, it can reduce the responsibility placed on the individual developer(s) from an ethical point of view.

---

### #3 Communication (Transparency)

**Motivation:** In practice, communication is a big part of being transparent with your stakeholders. Being transparent in communication can generate trust.

**What to Do:** Ask yourself:
- What is the goal of the system? Why is this particular system deployed in this specific area?
- What do you communicate about the system to its users and end-users? Is it enough for them to understand how the system works?
- If relevant to your system, do you somehow tell your (end-)users that they are interacting with an AI system and not with another human being?
- Do you collect user feedback? How is it used to change/improve the system?
- Are communication and transparency towards other audiences, such as the general public, relevant?

**Practical Example:** A medical system recommends diagnoses. How does the system communicate to doctors why it made a recommendation? How should the doctors know when to challenge the system? Does the system somehow change how patients and doctors interact?

---

### #7 Privacy and Data (Data)

**Motivation:** Privacy is a rising trend in the wake of various recent data misuse reveals. People are now increasingly conscious about handing out personal data. Similarly, regulations such as the GDPR now affect data collection.

**What to Do:** Ask yourself:
- What data are used by the system?
- Does the system use or collect personal data? Why? How is the personal data used?
- Do you clearly inform your (end-)users about any personal data collection? E.g., ask for consent, provide an opportunity to revoke it etc.
- Have you taken measures to enhance (end-user) privacy, such as encryption or anonymization?
- Who makes the decisions regarding data use and collection? Do you have organizational policies for it?

**Practical Example:** Rather than collecting and selling data, appealing to privacy can also be profitable. Regulations are making it increasingly difficult to collect lots of personal data for profit. Privacy can be an alternate selling point in today's climate.

---

### #8 Data Quality (Data)

**Motivation:** As AI are trained using data, the data used directly affects how the system operates. Both the nature and the quality and integrity of the data used has to align with the goals of the system.

**What to Do:** Ask yourself:
- What are good or poor quality data in the context of your system?
- How do you evaluate the quality and integrity of your own data? Are there alternative ways?
- If you utilize data from external sources, how do you control their quality?
- Did you align your system with relevant standards (for example ISO, IEEE) or widely adopted protocols for daily data management and governance?
- How can you tell if your data sets have been hacked or otherwise compromised?

**Practical Example:** In 2017, Amazon scrapped its recruitment AI because of bad data. They used past recruitment data to teach the AI. As they had mostly hired men, the AI began to consider women undesirable based on the data.

---

### #9 Access to Data (Data)

**Motivation:** Aside from carefully planning what data you collect and how, it is also important to plan how it can or will be used and by whom.

**What to Do:** Ask yourself:
- Who can access the users' data, and under what circumstances?
- How do you ensure that the people who access the data: 1) have a valid reason to do so; and 2) adhere to the regulations and policies related to the data?
- Do you keep logs of who accesses the data and when? Do the logs also tell why?
- Do you use existing data governance frameworks or protocols? Does your organization have its own?
- Who handles the data collection, storage, and use?

**Practical Example:** Third parties you give access to the data can misuse it. A prominent example of this is the case of Cambridge Analytica and Facebook, in which data from Facebook was used questionably. However, such incidents can also paint your organization in a bad light.

---

### #10 Human Agency (Agency & Oversight)

**Motivation:** People interacting with the system or using it should be able to understand it sufficiently. Users should be able to make informed decisions based on its suggestions, or to challenge its suggestions. AI systems should let humans make independent choices.

**What to Do:** Ask yourself:
- Does the system interact with decisions by human actors, i.e. end users (e.g. recommending users actions or decisions, or presenting options)?
- Does the system communicate to its (end-)users that a decision, content or outcome is the result of an algorithmic decision? Into how much detail does it go?
- In the system's use context, what tasks are done by the system and what tasks are done by humans?
- Have you taken measures to prevent overconfidence or overreliance on the system?

**Practical Example:** Assuming control is especially related to cyber-physical systems such as drones or other vehicles. For purely digital systems, the focus should be on supporting human decision-making instead of directing it.

---

### #11 Human Oversight (Agency & Oversight)

**Motivation:** AI systems should support human decision-making. They should not undermine human autonomy by making decisions for us, meaning they should be subject to human oversight.

**What to Do:** Ask yourself:
- Who can control the system and how? In what situations?
- What would be the appropriate level of human control for this particular system and its use cases?
- Related to the Safety and Security cards: how do you detect and respond if something goes wrong? Does the system then stop entirely, partially, or would control be delegated to a human? Why?

**Practical Example:** The autonomous nature of AI systems makes new vectors of attack possible. A white line drawn across a road can confuse a self-driving vehicle. What happened to Microsoft's Tay Twitter bot is another example of a new type of attack.

---

### #12 System Security (Safety & Security)

**Motivation:** While cybersecurity is important in any system, AI systems present new challenges. Cyber-physical systems can even cause fatalities in the hands of malicious actors.

**What to Do:** Ask yourself:
- Did you assess potential forms of attacks to which the system could be vulnerable? Did you consider ones that are unique or more relevant to AI systems?
- Did you consider different types of vulnerabilities, such as data pollution and physical infrastructure?
- Have you verified how your system behaves in unexpected situations and environments?
- Does your organization have cybersecurity personnel? Are they involved in the system?

---

### #13 System Safety (Safety & Security)

**Motivation:** AI systems exert notable influence on the physical world whether they are cyber-physical or not. Various risks and their consequences should be considered, thinking ahead to the operational life of the system.

**What to Do:** Ask yourself:
- What kind of risks does the system involve? What kind of damage could it cause?
- How do you measure and assess risks and safety?
- What fallback plans does your system have? Have they been tested?
- In what conditions do the fallback plans trigger? Are they automatic or do they require human input?
- Is there a plan to mitigate or manage technological errors, accidents, or malicious misuse? What if the systems provides wrong results, becomes unavailable, or provides societally unacceptable results?
- What liability and consumer protection laws apply to your system? Have you taken them into account?

**Practical Example:** AI systems can aid automating various organizational tasks, making it possible to reduce personnel. However, if a customer organization becomes reliant on your AI system to handle a portion of its operations, what happens if that AI stops functioning for even a few days? What could you do to alleviate the impact?

---

### #14 Accessibility (Fairness)

**Motivation:** Technology can be discriminating in various ways. Given the enormous impact AI systems can have, ensuring equal access to their positive impacts is ethically important.

**What to Do:** Ask yourself:
- Does the system consider a wide range of individual preferences and abilities? If not, why?
- Is the system usable by those with special needs or disabilities, those at risk of exclusion, or those using assistive technologies?
- Were people representing various groups somehow involved in the development of the system?
- How is the potential user audience taken into account?
- Is the team involved in building the system representative of your target user audience? Is it representative of the general population?
- Did you assess whether there could be (groups of) people

**Practical Example:** AI tends to benefit those who are already technologically capable, resulting in increased inequality. E.g. most of the images used in machine learning have been labeled by young white men.

---

### #15 Stakeholder Participation (Fairness)

**Motivation:** As AI systems have notable impacts, they stakeholders are also numerous. Though the system affects these various holders in various ways, they are often not involved in the development. Yet, e.g. when using a decision-making system, its users have to trust the system while also being critical of it.

**What to Do:** Turn to your stakeholder analysis (card #0):
- Which stakeholders are stakeholders in system development?
- How are the different stakeholders of the system involved in the development of the system? If they aren't, why?
- How do you inform your external and internal stakeholders of the system's development?

**Practical Example:** Often the people an AI system is used on are individuals who are simply objects for the system. For example, a medical system is developed for hospitals, used by doctors, but ultimately used on patients. Why not talk to the patients too?

---

### #16 Environmental Impact (Wellbeing)

**Motivation:** Past the general wellbeing implications, ecological consciousness is a current trend. Being ecological can be a selling point for your organization.

**What to Do:** Ask yourself:
- Did you assess the environmental impact of the system's development, deployment, and use? E.g., the type of energy used by the data centers.
- Did you consider the environmental impact when selecting specific technical solutions?
- Did you ensure measures to reduce the environmental impact of your system's life cycle?

**Practical Example:** If you are hosting on a third party cloud, try to ascertain the sustainability of the service provider's services. If you are using hardware, are you processing the data in each physical device of your own or are you processing it in the cloud?

---

### #17 Societal Effects (Wellbeing)

**Motivation:** The impacts a system has go beyond its userbase. A system may well affect negatively even those who do not use it nor wish to use it.

**What to Do:** Ask yourself:
- Did you assess the broader societal impact of the AI system's use beyond the individual (end-)users? Consider stakeholders who might be indirectly affected by the system.
- How will the systems affect society when in use?
- What kind of systemic effects could the system have?

**Practical Example:** Surveillance technology utilizing facial recognition AI has long-reaching impacts. People may wish to avoid areas that utilize such surveillance, negatively affecting businesses in said area. People may become stressed at the mere thought of such surveillance. Some may even emigrate as a result.

---

### #18 Auditability (Accountability)

**Motivation:** Regulations affecting AI and data may necessitate audits of systems in the future. Similarly, if the system causes damage, an audit might be requested. It is good to have mechanisms in place beforehand.

**What to Do:** Ask yourself:
- Is the system auditable?
- Can an audit be conducted independently?
- Is the system available for inspection?
- What mechanisms facilitate the system's auditability? How is traceability and logging of the system's processes and outcomes ensured?

**Practical Example:** In heavily regulated fields such as medicine, audits are typically required before a system can be utilized in the first place.

---

### #19 Ability to Redress (Accountability)

**Motivation:** Making sure people know they can be compensated in some way in the event something goes wrong with the system is important in generating trust. Such scenarios should be planned in advance to what extent possible.

**What to Do:** Ask yourself:
- What is your (developer organization) responsibility if the system causes damage or otherwise has a negative impact?
- In the event of negative impact, can the ones affected seek redress?
- How do you inform users and other third parties about opportunities for redress?

**Practical Example:** AI systems can inconvenience users in unforeseen, unpredictable ways. Depending on the situation, the company may or may not be legally responsible for the inconvenience. Nonetheless, by offering a digital platform for seeking redress, your company can seem more trustworthy while also offering additional value to your users.

---

### #20 Minimizing Negative Impacts (Accountability)

**Motivation:** Minimizing negative impacts of the system is financially important for any developer organization. Incidents are often costly.

**What to Do:**
- First, consider…
  - Is your stakeholder analysis is up-to-date (Card #0)
  - Have you discussed risks? (Card #13)
  - Have you discussed auditability? (Card #18)
  - Have you discussed redress issues? (Card #19)
- Are the people involved with the development of the system also involved with it during its operational life? If not, they may not feel as accountable.
- Are you aware of laws related to the system?
- Can users of the system somehow report vulnerabilities, risks, and other issues in the system?
- With whom have you discussed accountability and other ethical

---

## Card Themes

| | |
|---|---|
| Analyze | Data |
| Transparency | Agency & Oversight |
| Safety & Security | Wellbeing |
| Fairness | Accountability |

**JYU AI ethics Lab**

Ville Vakkuri JYU
ville.vakkuri@jyu.fi

Kai-Kristian Kemell
kai-kristian.o.kemell@jyu.fi

Pekka Abrahamsson JYU
pekka.abrahamsson@jyu.fi

Figure 1. ECCOLA - a Method for Implementing Ethically Aligned AI Systems

Each card in ECCOLA is split into three parts (figure 2): (1) motivation (i.e. why this is important), (2) what to do (to tackle this issue), and (3) a practical example of the topic (to make the issues more tangible). Each card also comes with a note-making space. As the cards are generally utilized as physical cards, the card is split into two with the left half of each card containing the textual contents and the right half containing white space for notes. This note-making space has been included to make using the cards more convenient in practice.

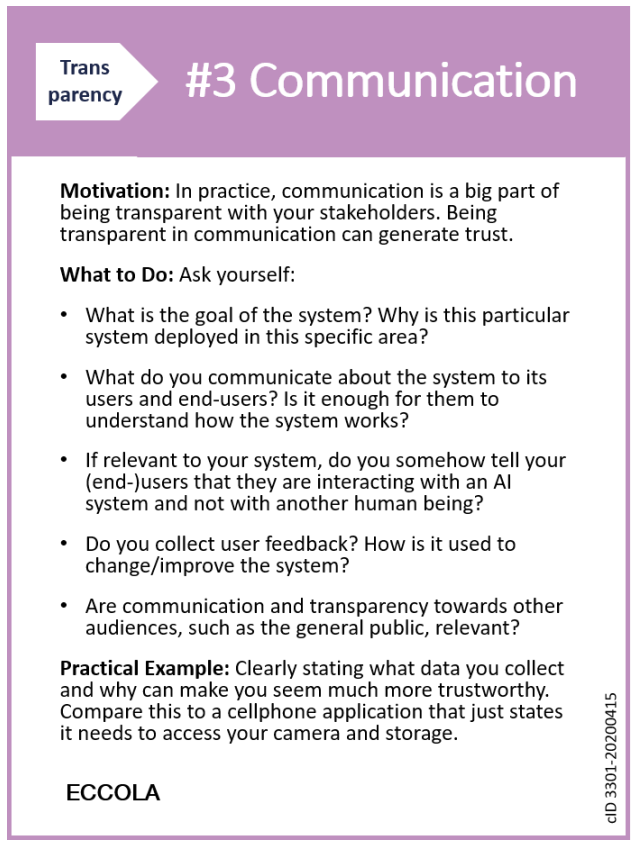

Figure 2. Card example from ECCOLA, Card #3 Communication

ECCOLA supports iterative development. During each iteration, the team is to choose which cards, or themes, are relevant for that particular iteration. ECCOLA is also method-agnostic, making it possible to utilize it with any existing or in-house SE method.

Depending on by whom ECCOLA is utilized, the tool has different goals. First, for product owners, the tool is intended to result in non-functional user stories involving ethics. Secondly, for a team of developers, the goal of ECCOLA is facilitating communication. By using the cards, the team will end up discussing ethical issues and making decisions based on the discussions. Finally, if utilized by a single developer, the goal of the method is raising awareness of ethical issues in AI. A single developer would instead dwell on these potential issues on their own while possibly looking further into the issues online for other points of view.

In developing ECCOLA, we have had three main goals for the method:

- To help create awareness of AI ethics and its importance
- To make an adaptable, modular method suitable for a wide variety of SE contexts, and
- To make ECCOLA suitable for agile development, and to also make ethics a part of agile development in general.

Next, we discuss how ECCOLA has been developed. It has been developed iteratively with multiple sets of data.

4. ECCOLA DEVELOPMENT PHASES AND DATA

ECCOLA has been developed iteratively through multiple phases (five, thus far). For this purpose, we have utilized the Cyclical Action Research method described by Susman and Evered [18] in developing it. In each phase, we have collected empirical data, based on which ECCOLA has been improved (figure 3).

The subsections of this section each cover one phase. In each subsection, we discuss what ECCOLA looked like at the time, how it was tested, and how it was changed based on the data. This process is also summarized in Table 1.

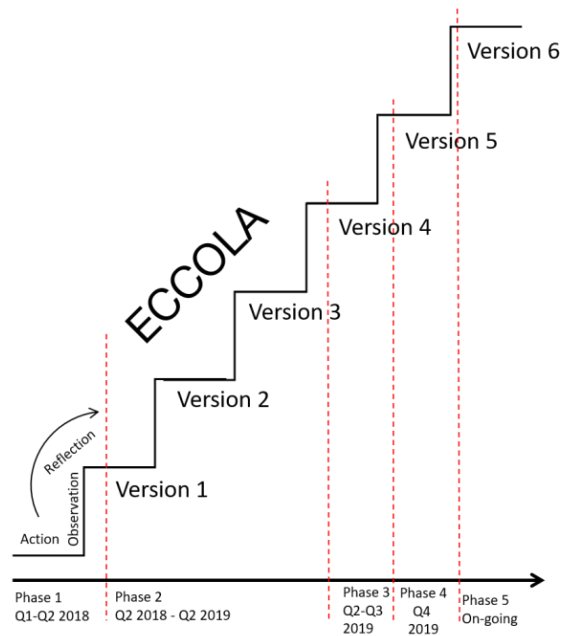

Figure 3. Cyclical Action Research process on ECCOLA. Including Cycle of Action, Observation, Reflection on each iteration

A. Phase 1 (Q1-Q2 2018)

In early 2018, prior to starting our work on ECCOLA, we searched for existing methods for AI ethics, ultimately finding none. Thus, we expanded our horizons and looked at ethical tools from other fields instead, to see if anything would seem applicable in the context of AI ethics as well. This led us to eventually test an existing ethical tool from the field of business ethics, the RESOLVEDD strategy [14], in the context of AI ethics. Our aim was to see if existing ethical tools, even if they were not specifically created for AI ethics, could be suitable for that context.

We conducted a scientific study on RESOLVEDD in the context of AI ethics. These findings have been published in-depth elsewhere (see Vakkuri & Kemell [15]). In short, we discovered that forcing developers to utilize RESOLVEDD did have some positive effects. Namely, it produced transparency in the development process, and the presence of an ethical tool made the developers aware of the potential importance of ethics, resulting in ethics-related discussions within the teams. However, the tool itself was not considered well-suited for the context by the respondents. Moreover, when forcing developers to utilize such a tool, the commitment towards it quickly vanished when the tool was no longer compulsive.

TABLE I. CYCLICAL ACTION RESEARCH PHASES

| Primary Phase | Version | Primary Background Theories | Study setting | Timing | Study Participants |
|---|---|---|---|---|---|
| 1 | N/A | RESOLVEDD, EAD, Essence | Class | Q1-Q2 2018 | 5 teams of 4-5 students |
| 2 | 1 | RESOLVEDD, EAD, Essence | Class | Q2 2018 - Q2 2019 | 27 teams of 3-5 students |
| 2 | 2 | RESOLVEDD, EAD, Essence | Class | Q2 2018 - Q2 2019 | 27 teams of 3-5 students |
| 2 | 3 | RESOLVEDD, EAD, Essence | Class | Q2 2018 - Q2 2019 | 27 teams of 3-5 students |
| 3 | 4 | EU AI HLEG, EAD | Blockchain Project | Q2-Q3 2019 | 2 sw development team members |
| 4 | 5 | EU AI HLEG, EAD | Conference Workshop | Q4 2019 | 8 researchers |
| 5 | 6 | EU AI HLEG, EAD | N/A | On-going | N/A |

B. Phase 2 (Q2 2018 - Q2 2019)

*1) Creating Version 1 (Q2 2018 - Q1 2019)*

Based on the results of the RESOLVEDD study, we began to develop a method of our own, ECCOLA, during the latter half of 2018. This initial version of the method was based on three primary theories: (1) RESOLVEDD strategy, (2) The Essence Theory of Software Engineering, and (3) The IEEE Ethically Aligned Design guidelines.

We utilized some of the general ideas of RESOLVEDD, which were deemed useful based on the data we collected. Namely, we took to RESOLVEDD for ideas on how to make the tool support iterative development. Additionally, we included some of the aspects of RESOLVEDD which were shown to support transparency of systems development (e.g. the idea of producing formal text documents while using the method).

We began to describe the method using the Essence language (see section 2.3). Methods described using Essence are visualized through cards, and thus, ECCOLA took on the form of a card deck as well. This also meant that we included the various elements of Essence into the cards. For example, we made some of the key AI ethics principles, namely transparency, accountability, and responsibility, into alphas in the context of Essence (i.e. measurable things to work on). The cards also included various activities that were to be performed in order to progress on these alphas, as well as patterns and other Essence elements.

The AI ethics contents of the method, at this stage, were based primarily on the IEEE Ethically Aligned Design guidelines [4]. We included key principles from the guidelines such as transparency and accountability, which have been prominent topics of discussion in AI ethics. Additionally, we utilized various research articles. For example, to expand on transparency, we utilized the studies of Dignum [9] and Ananny & Crawford [10], among others.

Much like how while using RESOLVEDD one produces text answering some questions posed by the tool, we incorporated the same idea of producing text while using ECCOLA into the initial version of the method. The theoretical background of this early version was based primarily on the IEEE EAD guidelines and the idea of the ART principles of AI Ethics [9].

*2) Testing Version 1 (Q1 2019)*

This first version of ECCOLA was tested in a large-scale project-based course on systems development at the University of Jyväskylä in the first quarter of 2019. In the course, 27 student teams of 4-5 students worked on a real-world case related to autonomous maritime. Each team was tasked with coming up with an innovation that would help make autonomous maritime possible. The teams were not required to actually develop these innovations into functional products, given the time and capability constraints in a course setting, but rather, to hone the ideas as far as they could in the context of the course. Some teams ultimately did produce technical demos, but this was not required. The results of these projects have been published in an educational book[4].

---

[4] https://jyx.jyu.fi/handle/123456789/63051

As any such innovation would involve AI directly or indirectly, given the autonomous maritime context, we chose to test ECCOLA by having these teams utilize it to reflect on the ethical issues their ideas might pose. The teams were introduced to ECCOLA during a course lecture and were handed a physical card deck. Each team was then told to utilize the card deck in whatever way they saw fit, while writing down notes on the cards as - or if - they used them. Additionally, unstructured interview data was collected from the teams through their weekly meetings with their assigned mentor and this feedback was taken into account in developing the method.

Prior to the course, the students had been tasked with reading a book on Essence, Software Engineering Essentialized [17], which explains the tool. Though the educational goal of this was elsewhere, this also served to make sure the students would not be overtly confused with this version of ECCOLA being described using the Essence language.

After the students had utilized the cards for a week, they were collected and the written notes on them analyzed. Based on this data, and the discussions the teams had had with their mentors in the weekly meetings, ECCOLA was improved as follows. First, alpha states were added to the alphas to make tracking progress on them easier. Secondly, practical examples were added to the cards to make the ethical issues on them more tangible to someone not versed in AI ethics. Thirdly, we improved the language on the cards, reducing academic jargon and focusing on practice. Finally, we removed the academic references that were initially present in each card. These were deemed to provide little value in raising awareness as none of the teams indicated having used them.

*3) Testing Version 2, (Q1 2019)*

This iteration took place during the same systems development course described in the preceding subsection. This iteration was carried out in the same manner as the previous one. The same student teams were tasked with utilizing the new version of ECCOLA again while writing down notes on them as they did. Additional data was again collected in the weekly mentor meetings. Overall, this was, in terms of time elapsed, a brief iteration carried out during the course.

After another week, ECCOLA was once more improved based on the data collected. We added a game sheet describing how the cards and the method should be used. This was done because it became clear that we had to teach the users of the method to use it as it lacked clear instructions. The cards were also numbered to make the method easier to grasp and to make it easier for the cards to refer to each other. To this end, we also improved the language on the cards, aiming to reduce academic jargon.

*4) Testing Version 3 (Q1 2019)*

As was the case with the previous two iterations in this phase, the third version of ECCOLA was tested in the systems development course in a similar manner. However, as this was towards the end of the course, there were no further iterations to be tested in the same setting. Thus, we took our time to analyze the feedback from all three versions, reflect on it, and study new publications in the area to improve the method.

This resulted in a lengthier creation process for the subsequent version. Based on the data and our reflection we made larger changes to the method. We discuss these in the following subsection.

*5) Creating Version 4 (Q2 2019)*

Data from phase 2 indicated that the method, though cumbersome to use, did help the teams implement AI ethics. The notes they had made on the cards showed that they had conducted ethical analyses successfully and changed their ideas based on their analyses. The AI ethics portion of the method thus worked. However, the method was not easy to use.

After the course had concluded, we had time to make larger improvements to the method based on the data. We opted to lessen the role of Essence in the method, forgoing the idea of using the Essence language to describe it. It seemed that Essence had made ECCOLA more confusing than it otherwise would have been, as in addition to learning the method, its users would have to learn the Essence notation and Essence in general. We stopped using the Essence elements in the cards and instead split the cards into different AI ethics themes. However, the general approach of using cards for the method seemed to work and thus this approach was kept.

The role of Essence in ECCOLA remains largely in relation to the idea of essentializing practices. This is described in the quote in section II C. ECCOLA aims to distill the essential parts of the AI ethics principles in the guidelines while making them more actionable through the card format.

Additionally, based on the data, the method seemed to be too heavy to use. ECCOLA was initially designed to be a linear process that was iteratively repeated. Its users, however, would be free to modify the process based on their development context and based on their use experience. Nonetheless, this approach was considered too rigid, and the respondents felt it was just another process tacked onto their other work processes. Moreover, the teams were using the method in a modular fashion, using individual cards as they deemed suitable, despite the instructions telling them to use it as a process.

We thus changed the approach, making the cards more stand-alone. In doing so, we wanted to make ECCOLA more modular by design, so that the users of the method could indeed choose which cards to utilize based on which ones they felt were relevant for their current situation. We felt that this would also make ECCOLA easier to use in conjunction with other methods.

During this time period, before the next empirical test, we also expanded the theoretical basis of the method. The initial version of the EU Guidelines for Trustworthy AI were published in early 2019, some aspects of which we chose to incorporate into ECCOLA. Other novel literature was also included to expand on theoretical basis of the method.

### C. Phase 3 (Q2-Q3 2019)

As the primary concern with the versions 1-3 had been the way ECCOLA was used as a method in practice rather than its AI ethical contents, we chose to focus on making a method that is easier and more practical to use. For this purpose, we made a spin-off of ECCOLA for the context of blockchain ethics. Many of the AI ethical themes such as transparency and data issues could be translated into this context, even if the contents of the cards had to be modified to be better suited for it. Additional blockchain specific issues were also added into these cards.

In this phase, ECCOLA was utilized in a real-world blockchain project by two of the project team members. Data was collected through observation and various unstructured interviews. The team was free to utilize the cards as they wished, and was encouraged to reflect on how the method would best suit their SE development method of choice. However, the team could also receive consultation from one of the researchers where needed on how to use the cards, as well for clarification on their contents, if needed. As a result, we gained a better understanding of how the method was utilized in practice (e.g., how many cards were used per iteration on average, which was 6) in a real-world SE context.

Notably, in this phase, ECCOLA was utilized in conjunction with existing SE methods, namely SCRUM. The feedback regarding the use of ECCOLA with another method was positive, lending support to the idea that ECCOLA does work as a modular method, especially with Agile methods. However, more testing is still needed in this regard in the future.

Based on the data gathered from the blockchain project, the main ECCOLA card deck was iteratively improved. The lessons learned from studying the use of the blockchain ethics version of ECCOLA were incorporated into ECCOLA. The data from this phase was primarily used to improve the contents of the cards by adding more contextual content (i.e. why these things are important) into each card. In this phase, the cards were also split into themes for clarity of presentation. Finally, stakeholder analysis was deemed to require more focus based on the data, and thus cards to support it were added.

### D. Phase 4 (Q4 2019)

After improving ECCOLA based on the lessons learned from the blockchain project, we presented ECCOLA at the 10th International Conference on Software Business, ICSOB2019 [5], in a workshop. In the workshop the participants utilized ECCOLA to discover potential ethical issues in a given, hypothetical AI development scenario. The participants of the workshop were split into two groups for the task.

The first group was tasked with developing an idea for an AI-based drone that would help farmers improve their harvests. The second group was tasked with developing an AI-based system that would filter and evaluate immigration applications. During the workshop, the groups worked on the ideas iteratively in timed sessions. Each group had a customer stakeholder that progressively presented them with more requirements at the end of each iteration. For every iteration, the groups were to select the cards they felt would be most relevant for the requirements of that iteration.

At the end of the workshop, verbal feedback from the participants was collected. This was done in the form of a discussion where the participants talked about their experiences with each other and between the two groups. These group interviews were recorded and later transcribed for analysis.

The feedback was then utilized to develop the current version of ECCOLA. The themes were color coded for further clarity of presentation. Additionally, we expanded the motivation and practical example portions of some of the cards to make them more stand-alone. E.g., in some cases, a user might have had to search online for more information on some past incident that was only mentioned by name.

### E. Phase 5 (On-going)

The development of ECCOLA continues. We argue that we have now reached a stage of maturity where ECCOLA can be brought forward to the scientific community. However, the method is not finalized and its development and testing continues in this iterative manner. The current version of ECCOLA, discussed in this paper, will again be tested and iteratively improved in the future (The most recent version is available at bit.ly/eccola-for-ai-ethics).

However, we feel that we have now reached a point of maturity where we wish to share the method with the scientific community. We discuss our reflections on the current state of ECCOLA in the next and final section of the paper in detail.

### 5. DISCUSSION AND CONCLUSIONS

In this paper, we have presented a method for implementing AI ethics: ECCOLA. ECCOLA is intended to help organizations develop more ethical AI systems by providing them with means of implementing AI ethics in a practical manner. ECCOLA has been developed iteratively using the Cyclical Action Research approach [18]. Though development on the method continues, we have reached a state of maturity where we want to share the method with the scientific community.

---

[5] https://icsob2019.wordpress.com/workshops/

The purpose of ECCOLA is to help us bridge the gap between research and practice in the area of AI ethics. Despite the increasing activity in the area, the academic discussion on AI ethics has not reached the industry [5]. Through ECCOLA, we have attempted to make some of the contents of the IEEE EAD guidelines [4] and the EU Trustworthy AI guidelines [3] actionable, alongside other research in the area.

In developing ECCOLA, we have had three main goals for the method:
- To help create awareness of AI ethics and its importance,
- To make an adaptable, modular method suitable for a wide variety of SE contexts, and
- To make ECCOLA suitable for agile development, and to also make ethics a part of agile development in general.

In relation to the first goal, there is currently no way of benchmarking what is, so to say, sufficiently ethical in the context of AI ethics. This is arguably a limitation for any such method in the context currently. Benchmarking ethics is difficult and thus it is equally difficult for a method to have a proven effect in a quantitative manner. Moreover, ethical issues are often context-specific and require situational reflection. This has been why we have instead chosen to focus on raising awareness and highlighting issues rather than trying to provide direct answers for them. Raising awareness has also been a goal of the IEEE EAD initiative [4]. Raising awareness is important as the area of AI ethics is new for the industry.

ECCOLA provides a starting point for implementing ethics in AI. Based on our lessons learned thus far, we argue that ECCOLA facilitates the implementation of AI ethics in two confirmable ways. First, ECCOLA raises awareness of AI ethics. It makes its users aware of various ethical issues and facilitates ethical discussion within the team. Secondly, ECCOLA produces transparency of systems development. In utilizing the method, a project team produces documentation of their ethical decision-making by means of e.g. making notes on the note-making space in the cards and non-functional requirements in product backlog. Transparency is one key issue in AI systems, both in terms of systems and in terms of systems development [9]. These documents, as we have done while testing the method, can also be analyzed to understand how the method was used, aside from seeking to understand the reasoning behind the ethical decisions that were made.

The second goal has been based on the method-agnostic philosophy of the Essence Theory of Software Engineering [12]. Industry organizations use a wide variety of methods, from out-of-the-box ones to, more commonly, tailored in-house ones [19]. ECCOLA is not intended to replace any of these. Rather, ECCOLA is intended as a modular tool that can be used in conjunction with any existing method. The use of ECCOLA in conjunction with agile methods and SE methods in general should still be further tested. For the time being, we received positive feedback relating to the modularity of ECCOLA when it was utilized in a project while using it in conjunction with SCRUM, an agile method (section IV C).

This, in turn, leads us to the third goal. As agile development is currently the trend, ECCOLA has been designed to be an iterative process from the get-go. However, during its iterative development, we noticed that a strict process was not a suitable approach due to being too heavy (section IV B). The users of the method opted out of adhering to the process and used the cards in a modular fashion despite the instructions. Now, ECCOLA is a modular tool by design. Being a card deck, this means that its users are able to select the cards they feel are relevant for each of their iterations, as opposed to having to go through the same process every time. Moreover, ECCOLA is intended to become a part of the agile development process in general. Ethics should not be merely an afterthought, but rather, a non-functional requirement, as well as a part of the user stories.

ECCOLA is a tool for developers and product owners. Ethics cannot be outsourced, nor can ethics be implemented by hiring an ethics expert [5]. AI ethics should be in the requirements, formulated in a manner also understood by the developers working on the system.

As governments and policy-makers have already begun to regulate AI systems in various ways (e.g. bans on facial recognition for surveillance purposes[6]), this trend is likely to only accelerate. With more and more regulations imposed on AI systems, organizations will need to tackle various AI ethics issues while developing their systems. This will consequently result in an increasing demand for methods in the area. While this will also inevitably result in the birth of various new methods, developed by companies, scholars, and standardization organizations alike, for the time being ECCOLA can serve as a starting point.

---

[6] https://www.bbc.com/news/technology-51148501